\documentclass{PoS}
\usepackage{amsmath,amssymb,amsopn,bm,slashed,epstopdf,grffile}

\newcommand{\dd}{\text{d}}
\newcommand{\ee}{\text{e}}
\newcommand{\ii}{\text{i}}

\newcommand{\beq}{\begin{equation}}
\newcommand{\eeq}{\end{equation}}
\newcommand{\bce}{\begin{center}}
\newcommand{\ece}{\end{center}}

\newcommand{\GeV}{\ensuremath{\mathrm{GeV}}}
\newcommand{\fm}{\ensuremath{\mathrm{fm}}}

\title{Non-equilibrium photon production arising from the chiral mass shift}

\ShortTitle{Non-equilibrium photon production arising from the chiral mass shift}

\author{Frank Michler\\
        University of Frankfurt\\
        E-mail: \email{michler@th.physik.uni-frankfurt.de}}
        
\author{Hendrik van Hees\\
        Frankfurt Institute for Advanced Studies\\
        E-mail: \email{hees@fias.uni-frankfurt.de}}
        
\author{Dennis D. Dietrich\\
        University of Frankfurt\\
        E-mail: \email{dietrich@th.physik.uni-frankfurt.de}}

\author{Stefan Leupold\\
        Uppsala Universitet\\
        E-mail: \email{stefan.leupold@physics.uu.se}}
        
\author{\speaker{Carsten Greiner}\\
        University of Frankfurt\\
        E-mail: \email{carsten.greiner@th.physik.uni-frankfurt.de}}
        
      \abstract{We investigate the photon emission arising from the
        shift of the quark/antiquark masses during the chiral phase
        transition in the early stage of ultrarelativistic heavy-ion
        collisions. As this mass shift leads to spontaneous creation of
        quark-antiquark pairs and thus contributes to the formation of
        the quark-gluon plasma, our investigations are relevant in the
        context of finite-lifetime effects on the photon emission from
        the latter. Earlier investigations on this topic by Boyanovsky
        et al.\ were plagued by a UV divergent contribution from the
        vacuum polarization and the non-integrability of the remaining
        contributions in the ultraviolet domain. In contrast to these
        investigations, we do not consider the photon number-density at finite
        times but for free asymptotic states obtained by an adiabatic
        switching of the electromagnetic interaction according to the
        Gell-Mann and Low theorem. This approach eliminates possible
        unphysical contributions from the vacuum polarization and
        renders the resulting photon spectra integrable in the
        ultraviolet domain. It is emphasized that the consideration of
        free asymptotic states is indeed crucial to obtain such
        physically reasonable results.}

\FullConference{International Winter Meeting on Nuclear Physics,\\
		21-25 January 2013\\
		Bormio, Italy }

\begin{document}

\section{Introduction and Motivation}

Ultrarelativistic heavy-ion collision experiments allow for studying
strongly interacting matter under extreme conditions. One main objective
of these experiments is the creation and exploration of the so-called
quark-gluon plasma (QGP), a state of matter of deconfined quarks and
gluons. The two basic features of the strong interaction, namely
confinement \cite{BS:1992} and asymptotic freedom
\cite{GW:1973,Pol:1973}, predict that this state is created at high
densities and temperatures
\cite{Shuryak:1978ij,Shuryak:1977ut,Yag:2005,Muller:2006ee,Friman:2011zz},
which occur during ultrarelativistic heavy-ion collisions.

The lifetime of the QGP created during a heavy-ion collision is in the
order of up to $10 \; \fm/c$. After that, it transforms into a gas of
hadrons. Thus, experiments cannot access the QGP directly, and the
investigations of its properties have to rely on experimental
signatures. One important category of these signatures are direct
photons as electromagnetic probes. As they only interact
electromagnetically, their mean free path is much larger than the
spatial extension of the QGP. Therefore, they leave the QGP almost
undisturbed once they have been produced and thus provide direct insight
into the early stage of the collision.

One important aspect in this context is that the QGP, as it occurs in a
heavy-ion collision, is not a static medium. It first thermalizes over a
finite interval of time, then keeps expanding and cooling down before it
hadronizes finally. This non-equilibrium dynamics has always been a
major motivation for investigations on non-equilibrium quantum field theory
\cite{Schw61,Bakshi:1962dv,Bakshi:1963bn,Kel64,Cra68,Danielewicz:1982kk,Chou:1984es,Landsman:1986uw,Greiner:1998vd,Berges:2001fi,Nahrgang:2011mg,Nahrgang:2011mv}.
Besides the role of possible memory effects during the time evolution
\cite{P1984305,Greiner:1994xm,Kohler:1995zz,Xu:1999aq,Juchem:2004cs,Juchem:2003bi,Schenke:2005ry,Schenke:2006uh,Michler:2009dy},
it is of particular interest how the finite lifetime QGP itself affects
the resulting photon spectra.

The first investigations on this topic by Boyanovsky et al.\
\cite{Wang:2000pv,Wang:2001xh,Boyanovsky:2003qm} have shown that the
finite lifetime of the QGP gives rise to the contribution to the photon
yield from first-order QED processes, i.e., processes of linear order in
the electromagnetic coupling constant, $\alpha_{e}$, which are
kinematically forbidden in thermal equilibrium. Furthermore, the
results from \cite{Wang:2000pv,Wang:2001xh,Boyanovsky:2003qm} suggested
that such non-equilibrium contributions dominate over leading-order
thermal contributions in the ultraviolet (UV) domain. The latter
contributions are linear both in the electromagnetic 
coupling constant, $\alpha_{e}$, and the strong coupling constant,
$\alpha_{s}$, and hence of overall second order.

On the other hand, however, the investigations in
\cite{Wang:2000pv,Wang:2001xh,Boyanovsky:2003qm} had been accompanied by
two artifacts. In the first place, the vacuum polarization was found to
lead to a divergent contribution to the photon number-density for given
photon energy. Furthermore, the remaining contributions to this quantity
did not decay fast enough with increasing photon energy for being
integrable in the UV domain. In particular, the total number
density and the total energy density of the emitted photons (after
subtracting the contribution from the vacuum polarization) were
logarithmically and linearly divergent, respectively.

Later on, the topic was also picked up by Fraga et al.\
\cite{Fraga:2003sn,Fraga:2004ur} where the ansatz used in
\cite{Wang:2000pv,Wang:2001xh,Boyanovsky:2003qm} was considered as
doubtful, as it came along with the mentioned problems. In particular,
the concerns raised in \cite{Fraga:2003sn,Fraga:2004ur} were the
following:
\begin{itemize}
\item In \cite{Wang:2000pv,Wang:2001xh,Boyanovsky:2003qm} the time at
  which the photons are observed has been kept finite. Either this
  corresponds to measuring photons that are not free asymptotic states
  or it corresponds to suddenly turning off the electromagnetic
  interaction at this point in time. Both cases are questionable.
\item The initial state in
  \cite{Wang:2000pv,Wang:2001xh,Boyanovsky:2003qm} has been specified
  for a system of thermalized quarks, antiquarks and gluons not containing any
  photons. On the other hand, however, a system of quarks and gluons,
  which undergoes electromagnetic interactions, necessarily contains
  photons. Hence, taking an initial state without any photons and
  without the Hamiltonian for electromagnetism corresponds to switching
  on the electromagnetic interaction at the initial time, which is
  questionable as well. It was shown in \cite{Arleo:2004gn} that the
  ansatz used in \cite{Wang:2000pv,Wang:2001xh,Boyanovsky:2003qm} is
  indeed equivalent to such a scenario.
\item The divergent contribution from the vacuum polarization is
  unphysical and thus needs to be renormalized. Nevertheless, the
  renormalization procedure presented in \cite{Boyanovsky:2003qm} is not
  coherent since no derivation of the photon yield with rescaled field
  operators has been presented in \cite{Boyanovsky:2003qm}.
\end{itemize}
The authors of \cite{Fraga:2003sn,Fraga:2004ur} did, however, not
provide an alternative approach for how to handle the mentioned problems
in a consistent manner. Solely in \cite{Arleo:2004gn} it was indicated
that the question of finite-lifetime effects could be addressed within
the 2PI (two-particle irreducible) approach even though the conservation
of gauge invariance remains challenging. Later on Boyanovsky et al.
insisted on their approach \cite{Boyanovsky:2003rw} and objected to the
arguments by \cite{Fraga:2003sn,Fraga:2004ur} as follows:
\begin{itemize}
\item Non-equilibrium quantum field theory is an initial-value problem.
  This means that the density matrix of the considered system is first
  specified at some initial time and then propagated to a later time by
  the time-evolution operator. For that reason, the Hamiltonian is not modified
  by introducing a time-dependent artificial coupling as it would be the
  case for a `switching on' and a later `switching off' of the
  electromagnetic interaction.
\item The quark-gluon plasma, as it occurs in a heavy-ion collision, has
  a lifetime of only a few $\fm/c$.  Therefore, taking the time to
  infinity is unphysical as this limit requires the inclusion of
  non-perturbative phase-transition effects on the photon production.
\item The renormalization technique of \cite{Boyanovsky:2003rw} provides
  a rescaling of the photon field operators such that the photon number
  operator actually counts asymptotic photon states with amplitude one.
\end{itemize}

During this debate, however, the original problems encountered in
\cite{Wang:2000pv,Wang:2001xh,Boyanovsky:2003qm} remained open. In order
to resolve them in a satisfactory manner, we have followed two different
approaches.

In the first approach \cite{Michler:2009hi}, we have modeled the finite
lifetime of the (thermalized) quark-gluon plasma during a heavy-ion
collision by time-dependent occupation numbers in the photon
self-energy. By means of this procedure, we have been able to
renormalize the divergent contribution from the vacuum polarization in a
consistent manner. However, it has not been possible to get the problem
with the UV behavior fully under control.

We have expected in the first place that this remaining shortcoming
results from a violation of the Ward-Takahashi identities. For that
reason, we have also considered a conceptionally different scenario
\cite{Michler:2012mg} where the production of quark-antiquark pairs
together with photons results from a change in the quark-antiquark
mass. Such a change occurs during the chiral phase transition in the
very early stage of a heavy-ion collision. It has been shown in
\cite{Greiner:1995ac,Greiner:1996wz} that it leads to the spontaneous
and non-perturbative pair production of quarks and antiquarks, which
effectively contributes to the formation of the QGP. We have
investigated the photon production arising from this pair-creation
process. As this photon production is effectively induced by the change
of the quark/antiquark mass and thus by the chiral phase transition, it
is from now on referred to as chiral photon production.

In contrast to \cite{Michler:2009hi}, such a scenario has the crucial
advantage that it allows for a first-principle description by
introducing a Yukawa-like source term in the QED Lagrangian. This source
term couples the quarks and antiquarks to a time-dependent, scalar
background field. In this way, they obtain a time-dependent mass, which
conserves the Ward-Takahashi identities. As in \cite{Michler:2009hi}, we
have restricted ourselves to first-order QED processes. They are
kinematically possible since the quarks and antiquarks obtain additional
energy by the coupling to the background field. To the contrary, the
coupling to the background field is resummed to all orders as to
properly take into account the non-perturbative nature of the pair-creation 
process.

In this context, there is another crucial difference to the approaches
in \cite{Wang:2000pv,Wang:2001xh,Boyanovsky:2003qm,Michler:2009hi}:
There the photon number-density has been considered at finite times. In the
course of our investigations \cite{Michler:2012mg}, however, we have
shown that the photon number-density has to be extracted in the limit
$t\rightarrow\infty$ for free asymptotic states, which are the only
observable ones since they reach the detectors. Such states have been
obtained by introducing an adiabatic switching of the electromagnetic
interaction according to the Gell-Mann and Low theorem. The photon
number-density is then considered in the asymptotic time limit and the
switching parameter is taken to zero at the very end of our calculation.

We shall demonstrate that this procedure eliminates a possible
unphysical vacuum contribution and, furthermore, leads to photon spectra
being integrable in the ultraviolet domain if the time evolution of the
quark/antiquark mass is modeled in a physically reasonable manner. It is
also emphasized that the consideration of the photon number-density for
free asymptotic states is indeed essential to obtain such physically
reasonable results.

\section{Model description}

Before we turn to our first-principle description on chiral photon
production, we first provide a short insertion on our model description
\cite{Michler:2009hi}, where we have made an ansatz on the two-time
dependence of the photon self-energy. For this purpose, we have taken
into account that the vacuum contribution to the photon self-energy is always
persistent, whereas the medium contribution only occurs as long as the
QGP is actually present.  These two aspects had been disregarded in
\cite{Wang:2000pv,Wang:2001xh,Boyanovsky:2003qm}. This time dependence
has been in implemented by introducing time-dependent quark/antiquark
occupation numbers
\begin{equation}
 \label{eq:2:time_dep}
 n_{\text{F}}(E) \rightarrow n_{\text{F}}(E,t) = f(t)n_{\text{F}}(E) \ ,
\end{equation}
in the photon self-energy. Here $n_{\text{F}}(E)$ denotes the
Fermi-Dirac distribution at the given temperature, $T$, and $f(t)$ is
the function modeling its time evolution. This function changes
monotonously from zero for $t \rightarrow -\infty$ (vacuum) to one for
$t\rightarrow\infty$ (QGP fully persistent) over an assumed formation
time interval, $\tau$.  As in
\cite{Wang:2000pv,Wang:2001xh,Boyanovsky:2003qm}, we have considered the
photon number-density at finite times.

\begin{figure}[htb]
 \begin{center}
  \includegraphics[height=5.0cm]{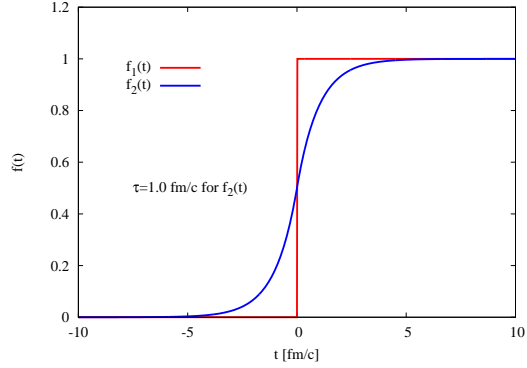}
  \caption{The time evolution of the QGP is modeled by different switching functions, $f_{i}(t)$.}
  \label{fig:2:switching}
 \end{center}
\end{figure}

By means of this procedure, we have been able to renormalize the
divergent contribution from the vacuum polarization in a consistent
manner. For our numerical investigations on the remaining medium
contributions, we have compared the resulting photon spectra for
different switching functions, $f_{i}(t)$, which are depicted in Figure
\ref{fig:2:switching}. As one can see, $f_{1}(t)$ describes an
instantaneous formation of the QGP at $t=0$, whereas $f_{2}(t)$
describes a formation over a finite interval of time, $\tau$, for which
we have assumed $\tau=1.0 \; \fm/c$.

\begin{figure}[htb]
 \begin{center}
  \includegraphics[height=5.0cm]{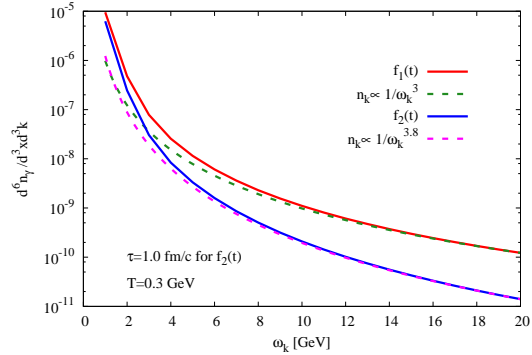}
  \caption{Comparison of the photon spectra for different switching
    functions. For $f_{1}(t)$ describing an instantaneous formation, the
    photon spectrum decays as $1/\omega^{3}_{\vec{k}}$ and exhibits only
    a slightly steeper decay $\propto1/\omega^{3.8}_{\vec{k}}$ for
    $f_{2}(t)$. $\omega_{\vec{k}}=|\vec{k}|$ denotes the photon energy
    for given photon three-momentum, $\vec{k}$.}
  \label{fig:2:photspec}
 \end{center}
\end{figure}

It has been our initial hope that in addition to the renormalization of
the vacuum contribution our model description would also lead to
UV integrable photon spectra if the the formation of the QGP is assumed
to occur over a finite interval of time. Such an assumption is
reasonable from the phenomenological point of view. A comparison of the
photon spectra for the different switching functions, which is done in
Figure \ref{fig:2:photspec}, does, however, show that this is only
partly the case.

For the case of an instantaneous formation the photon number-density
scales as $1/\omega^{3}_{\vec{k}}$ (with $\omega_{\vec{k}}=|\vec{k}|$
denoting the photon energy for given photon three-momentum, $\vec{k}$)
in the UV domain. This implies that the total number density and the
total energy density of the radiated photons are logarithmically and
linearly divergent, respectively.  This artifact is only partly resolved
if we turn to a formation over a finite interval of time describing a
physically more realistic scenario. In this case, the UV scaling
behavior of the photon number-density is only suppressed to
$1/\omega^{3.8}_{\vec{k}}$ such that only the total number density is
rendered UV finite whereas the total energy density remains UV
divergent. For our numerical investigations, we have chosen $f_{2}(t)$
to be continuously differentiable once but one can show that this
artifact persists for arbitrarily smooth switching functions.

We have assumed in the first place that the still problematic UV
behavior arises from a violation of the Ward-Takahashi identities for
the photon self-energy within our model description
\cite{Michler:2009hi}. Therefore, we aimed to find a new approach in
which these identities are conserved. For that reason, we have
eventually studied chiral photon production as this scenario allows for
an accordant first-principle description.

\section{Chiral photon production}
We model the change of the quark/antiquark mass during the chiral phase
transition by introducing a Yukawa-like source term in the QED
Lagrangian
\begin{equation}
 \label{eq:3:lagrangian_source}
 \mathcal{\hat{L}}(x) = \mathcal{\hat{L}}_{\text{QED}}(x)-g\phi(t)\hat{\bar{\psi}}(x)\hat{\psi}(x) \ . 
\end{equation}
The source field, $\phi(t)$, is assumed to be classical and
time-dependent only, which effectively assigns the quarks and antiquark
a time-dependent mass
\begin{equation}
 \label{eq:3:mass_timedep}
 m(t)=m_{c}+g\phi(t) \ .
\end{equation}
Since such an ansatz keeps the Lagrangian gauge invariant it conserves
the Ward-Takahashi identities of QED, which we will demonstrate
explicitly below for the photon self-energy.

The change of the quark/antiquark mass from its constituent value,
$m_{c}$, to its bare value, $m_{b}$ leads to the spontaneous and
non-perturbative pair creation of quarks and antiquarks. We investigate
the photon emission resulting from this pair-creation process. Thereby,
we assume that our system does not contain any quarks/antiquarks or
photons initially. The initial density matrix is hence given by
\begin{equation}
 \label{eq:3:ini_den}
 \hat{\rho}(t_{0}) = \left|0_{q\bar{q}}\right\rangle\otimes\left|0_{\gamma}\right\rangle \ .
\end{equation}
Here $\left|0_{q\bar{q}}\right\rangle$ and
$\left|0_{\gamma}\right\rangle$ denote the vacuum states of the
fermionic and the photonic sector, respectively. As the quark/antiquark mass
changes with time, it is important to point out that the former is
defined with respect to the initial constituent mass, $m_{c}$. The
initial time, $t_{0}$, is chosen from the domain where the
quark/antiquark mass is still at this value, i.e., $t_{0}<t^{'}_{0}$
with $t^{'}_{0}$ denoting the time at which the change of the
quark/antiquark mass begins. In the case of parameterizations, $m(t)$,
where the time derivative, $\dot{m}(t)$, has a non-compact support, it
is sufficient to ensure that one has $|g\phi(t)|\ll m_{c}$ for $t\le
t^{'}_{0}$.

\begin{figure}[htb]
 \begin{center}
  \includegraphics[height=5.0cm]{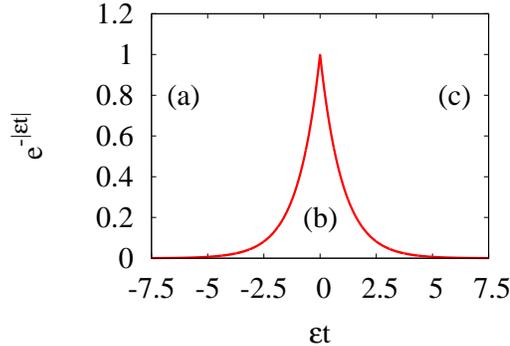}
  \caption{Schematic representation of our asymptotic description. The
    initial state is specified for $t_{0}\rightarrow -\infty$ (a). From
    the adiabatic switching of the electromagnetic interaction, one then
    has interacting fields around $t=0$ (b) which evolve into free
    fields in the asymptotic time limit $t\rightarrow\infty$ (c), where
    the photon number-density is considered. Only at the very end of our
    calculation, the switching parameter, $\varepsilon$, is taken to
    zero.}
  \label{fig:3:adiabatic}
 \end{center}
\end{figure}

Since the quark mass is time-dependent only and our system is initially
given by the vacuum state (\ref{eq:3:ini_den}), it is spatially
homogeneous. For such a system, the photon number-density is formally
given by
\begin{equation}
 \label{eq:3:phot_num_den}
 2\omega_{\vec{k}}\frac{\dd^{6}n_{\gamma}(t)}{\dd^{3}x\dd^{3}k} =
 \frac{1}{(2\pi)^{3}V}\sum_{\lambda=\perp}\left\langle
   \hat{a}^{\dagger}(\vec{k},\lambda,t)\hat{a}(\vec{k},\lambda,t)\right\rangle\ . 
\end{equation}
The sum runs over all physical (transverse) polarizations and the
average is taken with respect to the initial density matrix
(\ref{eq:3:ini_den}). Moreover, $\hat{a}(\vec{k},\lambda,t)$ together
with its Hermitian conjugate corresponds to the expansion coefficients
in the plane-wave decomposition of the photon-field operator,
$\hat{A}_{\mu}(\vec{x},t)$, i.e.,
\begin{equation}
 \hat{A}_{\mu}(x) = \sum_{\lambda}\int\frac{\dd^{3}k}{(2\pi)^{3}}\frac{1}{\sqrt{2\omega_{\vec{k}}}}
                      \left[
                       \epsilon_{\mu}(\vec{k},\lambda)\hat{a}(\vec{k},\lambda,t)\ee^{\ii\vec{k}\cdot\vec{x}}+
                       \epsilon^{*}_{\mu}(\vec{k},\lambda)\hat{a}^{\dagger}(\vec{k},\lambda,t)\ee^{-\ii\vec{k}\cdot\vec{x}}
                      \right] \ .
\end{equation}
Expression (\ref{eq:3:phot_num_den}) for the photon number-density has
also been used in \cite{Wang:2000pv,Wang:2001xh,Boyanovsky:2003qm} with
the initial density matrix, $\hat{\rho}(t_{0})$, being instead specified
for a thermalized QGP not containing any photons. However, before one
can start with any further evaluation and numerical analysis of
(\ref{eq:3:phot_num_den}), one has to clarify under which circumstances
$\hat{a}(\vec{k},\lambda,t)$ and its Hermitian conjugate actually allow
for an interpretation as a single-photon operator. Such an
interpretation is not given in general since $\hat{A}_{\mu}(\vec{x},t)$
describes an interacting electromagnetic field. It is, however, possible
in the asymptotic time limit $t\rightarrow\infty$ for free asymptotic
fields. Such fields are obtained by introducing an adiabatic switching
of the electromagnetic interaction according to the Gell-Mann and Low
Theorem, i.e.,
\begin{equation}
 \label{eq:3:adiabatic}
 \hat{H}_{\text{EM}}(t) = e\int\dd^{3}x\mbox{ }\hat{\bar{\psi}}(x)\gamma_{\mu}\hat{\psi}(x)\hat{A}^{\mu}(x)
                          \rightarrow
                          \ee^{-\varepsilon|t|}\hat{H}_{\text{EM}}(t) \
                          , \quad\text{with}\quad \varepsilon > 0 \ .
\end{equation}
In order to obtain physically well-defined results for the photon
number-density, we have to specify our initial state for
$t_{0}\rightarrow-\infty$ and consider (\ref{eq:3:phot_num_den}) in the
limit $t\rightarrow\infty$ for free asymptotic states. This procedure is
illustrated in Figure \ref{fig:3:adiabatic}.  Since the introduction of
an adiabatic switching of the electromagnetic interaction per se is an
artificial procedure we have to take $\varepsilon\rightarrow0$ at the
very end of our calculation.

\begin{figure}[htb]
 \begin{center}
  \includegraphics[height=5.0cm]{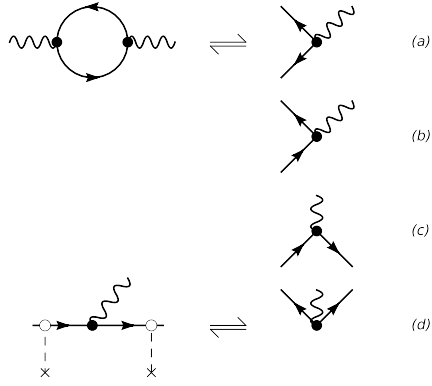}
  \caption{The photon self-energy,
    $\ii\Pi^{<}_{\text{T}}(\vec{k},t_{1},t_{2})$, is given by the
    one-loop approximation including the first-order QED
    processes. These processes are quark Bremsstrahlung (a), antiquark
    Bremsstrahlung (b), quark-antiquark pair annihilation into a single
    photon (c), and the spontaneous creation of a quark-antiquark pair
    together with a photon out of the vacuum (d). These processes are
    possible because of the coupling of the quarks and antiquarks to the
    background field, $\phi(t)$. This coupling is depicted by the open
    circles as to distinguish it from the electromagnetic coupling
    depicted by the full circles.}
  \label{fig:3:one_loop_split}
 \end{center}
\end{figure}

As the electromagnetic coupling is small, we evaluate
(\ref{eq:3:phot_num_den}) to first order in $\alpha_{e}$, but keep all
orders in $g$. The latter is achieved by constructing an
interaction-picture representation that incorporates the source term
\cite{Michler:2012mg}. Together with (\ref{eq:3:adiabatic}), this
procedure leads to
\begin{equation}
 \label{eq:3:photnum_prelim}
 2\omega_{\vec{k}}\frac{\dd^{6}n^{\varepsilon}_{\gamma}(t)}{\dd^{3}x\dd^{3}k}
   = \frac{1}{(2\pi)^{3}}\int_{-\infty}^{t}\underline{\dd t_{1}}\int_{-\infty}^{t}\underline{\dd t_{2}}
     \ii\Pi^{<}_{\text{T}}(\vec{k},t_{1},t_{2})\ee^{\ii\omega_{k}(t_{1}-t_{2})} \ ,
\end{equation}
with the underlining denoting that
$$
\int_{-\infty}^{t}\underline{\dd u} = \int_{-\infty}^{t}\dd u\mbox{ }\ee^{-\varepsilon|u|} \ . 
$$
Moreover, $\ii\Pi^{<}_{\text{T}}(\vec{k},t_{1},t_{2})$ describes the
transverse part of the photon self-energy, i.e.,
\begin{equation}
 \label{eq:3:pse_trans}
 \ii\Pi^{<}_{\text{T}}(\vec{k},t_{1},t_{2})=\gamma^{\mu\nu}(\vec{k})\ii\Pi^{<}_{\nu\mu}(\vec{k},t_{1},t_{2}) \ .
\end{equation}
$\gamma^{\mu\nu}(\vec{k})$ is the photon polarization tensor reading
\begin{equation}
 \label{eq:3:polten}
 \gamma^{\mu\nu}(\vec{k}) = \sum_{\lambda=\perp}\epsilon^{\mu,*}(\vec{k},\lambda)\epsilon^{\nu}(\vec{k},\lambda)
 = \begin{cases}
   -\eta^{\mu\nu}-\frac{k^{\mu}k^{\nu}}{\omega^{2}_{\vec{k}}} &
   \text{for}\quad\mu,\nu\in\left\lbrace 1,2,3\right\rbrace \\ 
   0 & \text{otherwise}
 \end{cases} \ ,
\end{equation}
with $\eta^{\mu\nu}=\text{diag}\left\lbrace1,-1,-1,-1\right\rbrace$. The
photon self-energy, $\ii\Pi^{<}_{\mu\nu}(\vec{k},t_{1},t_{2})$, is in
turn given by the one-loop approximation
\begin{equation}
 \label{eq:3:pse}
 \ii\Pi^{<}_{\mu\nu}(\vec{k},t_{1},t_{2}) = e^{2}\int\frac{\dd^{3}p}{(2\pi)^{3}}\mbox{Tr}
                                            \left\lbrace
                                             \gamma_{\mu}S^{<}_{\text{F}}(\vec{p}+\vec{k},t_{1},t_{2})
                                             \gamma_{\nu}S^{>}_{\text{F}}(\vec{p},t_{2},t_{1})
                                            \right\rbrace \ ,
\end{equation}
with the propagators fulfilling the equations of motion
\begin{subequations}
 \label{eq:3:eom_prop}
 \begin{eqnarray}
  \left[\ii\gamma^{0}\partial_{t_{1}}+\gamma^{i}p_{i}-m(t_{1})\right]S^{\lessgtr}_{\text{F}}(\vec{p},t_1,t_2) & = & 0 \ , \\ 
  \left[\ii\gamma^{0}\partial_{t_{2}}-\gamma^{i}p_{i}+m(t_{2})\right]S^{\lessgtr}_{\text{F}}(\vec{p},t_1,t_2) & = & 0 \ .
 \end{eqnarray}
\end{subequations}
It follows from these equations that the photon self-energy (\ref{eq:3:pse}) fulfills the Ward-Takahashi identities
\begin{subequations}
 \label{eq:3:wti}
 \begin{eqnarray}
  \partial_{t_{1}}\ii\Pi^{<}_{0\mu}(\vec{k},t_{1},t_{2})-\ii k^{j}\ii\Pi^{<}_{j\mu}(\vec{k},t_{1},t_{2}) = 0 \ , \label{eq:3:wti_t1} \\
  \partial_{t_{2}}\ii\Pi^{<}_{\mu0}(\vec{k},t_{1},t_{2})+\ii k^{j}\ii\Pi^{<}_{\mu j}(\vec{k},t_{1},t_{2}) = 0 \ . \label{eq:3:wti_t2}
 \end{eqnarray}
\end{subequations}
The one-loop approximation for the photon self-energy includes the
processes of first order in $\alpha_{e}$, which is shown in Figure
\ref{fig:3:one_loop_split}. In particular, these processes are
(one-body) quark/antiquark Bremsstrahlung, quark-antiquark pair
annihilation into a single photon and the spontaneous creation of a
quark-antiquark pair together with a photon out of the vacuum. They are
kinematically possible since the quarks and antiquarks obtain additional
energy by the coupling to the background field, $\phi(t)$.

As the above diagrammatic arguments indicate, the photon number-density
(\ref{eq:3:photnum_prelim}) can be written as the absolute square of a
first-order QED transition amplitude and, as a consequence, is positive
semidefinite. To demonstrate this explicitly, we first undo the contraction
(\ref{eq:3:pse_trans}) in (\ref{eq:3:photnum_prelim}), which leads to
\begin{equation}
  \label{eq:3:photnum_uncont}
  2\omega_{\vec{k}}\frac{\dd^{6}n^{\varepsilon}_{\gamma}(t)}{\dd^{3}x\dd^{3}k}
   =
   \frac{\gamma^{\mu\nu}(\vec{k})}{(2\pi)^{3}}\int_{-\infty}^{t}\underline{\dd
     t_{1}}\int_{-\infty}^{t}\underline{\dd t_{2}} 
     \ii\Pi^{<}_{\nu\mu}(\vec{k},t_{1},t_{2})\ee^{\ii\omega_{k}(t_{1}-t_{2})} \ .
\end{equation}
As the next step, we expand the fermion propagators in terms of
positive- and negative-energy wavefunctions,
$\psi_{\vec{p},s\uparrow\downarrow}(t)$ for given momentum, $\vec{p}$,
and spin, $s$, i.e.,
\begin{subequations}
 \label{eq:3:prop_decom}
 \begin{eqnarray}
  S^{>}_{\text{F}}(\vec{p},t_{1},t_{2}) & = &
  -\ii\sum_{s}\psi_{\vec{p},s,\uparrow}(t_{1})\bar{\psi}_{\vec{p},s,\uparrow}(t_{2})
  \ , \\ 
  S^{<}_{\text{F}}(\vec{p},t_{1},t_{2}) & = &
  \ii\sum_{s}\psi_{\vec{p},s,\downarrow}(t_{1})\bar{\psi}_{\vec{p},s,\downarrow}(t_{2})
\ .
 \end{eqnarray}
\end{subequations}
These wavefunctions fulfill the equation of motion
\begin{equation}
 \left[\ii\gamma^{0}\partial_{t}+\gamma^{i}p_{i}-m(t)\right]\psi_{\vec{p},s,\uparrow\downarrow}(t) \ ,
\end{equation}
with the initial conditions
\begin{subequations}
 \begin{align}
  \psi_{\vec{p},s,\uparrow}(t)   \rightarrow \psi^{c}_{\vec{p},s,\uparrow}(t)   & \quad\mbox{for}\quad t\le t^{'}_{0} \ , \\ 
  \psi_{\vec{p},s,\downarrow}(t) \rightarrow \psi^{c}_{\vec{p},s,\downarrow}(t) & \quad\mbox{for}\quad t\le t^{'}_{0} \ .
 \end{align}
\end{subequations}
Here $\psi^{c}_{\vec{p},s,\uparrow\downarrow}(t)$ denote the positive- and negative-energy spinors with respect to the initial, 
constituent mass, $m_{c}$. Upon insertion of (\ref{eq:3:pse_trans}) and relations (\ref{eq:3:prop_decom}) into (\ref{eq:3:photnum_uncont}), we 
can finally rewrite the photon number-density as
\begin{equation}
 2\omega_{\vec{k}}\frac{\dd^{6}n^{\varepsilon}_{\gamma}(t)}{\dd^{3}x\dd^{3}k}
 = e^{2}\sum_{r,s,{\lambda}}\frac{\dd^{3}p}{(2\pi)^{3}}\left|\int_{-\infty}^{t}\underline{\dd u}\mbox{ }
   \epsilon^{\mu}(\vec{k},\lambda)\cdot\bar{\psi}_{\vec{p},r,\uparrow}(u)\gamma_{\mu}
   \psi_{\vec{p}+\vec{k},s,\downarrow}(u)\ee^{\ii\omega_{\vec{k}}u}\right|^{2} \ .
\end{equation}
This absolute-square representation ensures that
(\ref{eq:3:photnum_prelim}) cannot acquire unphysical negative
values. The physical photon number-density is extracted from
(\ref{eq:3:photnum_prelim}) by taking the successive limits
$t\rightarrow\infty$ and $\varepsilon\rightarrow0$ after the time
integrations have been performed, i.e.,
\begin{equation}
  \label{eq:3:phot_num_den_phys}
  2\omega_{\vec{k}}\frac{\dd^{6}n_{\gamma}}{\dd^{3}x\dd^{3}k}
   = \lim_{\varepsilon\rightarrow0}\frac{1}{(2\pi)^{3}}\int_{-\infty}^{\infty}\underline{\dd t_{1}}\int_{-\infty}^{\infty}\underline{\dd t_{2}}
     \ii\Pi^{<}_{\text{T}}(\vec{k},t_{1},t_{2})\ee^{\ii\omega_{k}(t_{1}-t_{2})} \ .
\end{equation}
$\ii\Pi^{<}_{\text{T}}(\vec{k},t_{1},t_{2})$ reduces to the vacuum
polarization if both time arguments are taken from the domain where the
quark/antiquark mass is still at its initial constituent value, $m_{c}$,
\begin{eqnarray}
 \label{eq:3:vacpol}
 \ii\Pi^{<}_{\text{T},0}(\vec{k},t_{1},t_{2}) & = & \ii\Pi^{<}_{\text{T},0}(\vec{k},t_{1}-t_{2}) \nonumber \\
 & = & 2e^{2}\int\frac{\dd^{3}p}{(2\pi)^{3}}
 \left\lbrace
   1+\frac{px(px+k)+m^{2}_{c}}{E^{c}_{\vec{p}}E^{c}_{\vec{p}+\vec{k}}}
 \right\rbrace
 \ee^{\ii\left(E^{c}_{\vec{p}+\vec{k}}+E^{c}_{\vec{p}}\right)(t_{1}-t_{2})} \ .
\end{eqnarray}
From the mass-shift effects,
$\ii\Pi^{<}_{\text{T}}(\vec{k},t_{1},t_{2})$ will acquire an additional
non-stationary contribution, i.e.,
\begin{equation}
 \ii\Pi^{<}_{\text{T}}(\vec{k},t_{1},t_{2}) =
 \ii\Pi^{<}_{\text{T},0}(\vec{k},t_{1}-t_{2})+\ii\Delta\Pi^{<}_{\text{T}}(\vec{k},t_{1},t_{2})
 \ ,
\end{equation}
depending on both time arguments explicitly. When performing the
successive limits $t\rightarrow\infty$ and $\varepsilon\rightarrow0$ the
contribution from the vacuum polarization (\ref{eq:3:vacpol}) is
eliminated. This can be seen by inserting (\ref{eq:3:vacpol}) into
(\ref{eq:3:phot_num_den_phys}), which leads to
\begin{align}
  \left.\omega_{\vec{k}}\frac{\dd^{6}n_{\gamma}}{\dd^{3}x\dd^{3}k}\right|_{\text{VAC}}
  = & \lim_{\varepsilon\rightarrow0}
  \frac{e^{2}}{(2\pi)^{3}}\int\frac{\dd^{3}p}{(2\pi)^{3}} \left\lbrace
    1+\frac{px(px+\omega_{\vec{k}})+m^{2}_{c}}{E^{c}_{\vec{p}}E^{c}_{\vec{p}+\vec{k}}}
  \right\rbrace \left[
    \frac{2\varepsilon}{\varepsilon^{2}+\left(E^{c}_{\vec{p}+\vec{k}}+E^{c}_{\vec{p}}+\omega_{\vec{k}}\right)^{2}}
  \right]^{2} \nonumber \\
  < & \lim_{\varepsilon\rightarrow0}
  \frac{4e^{2}}{(2\pi)^{3}}\int\frac{\dd^{3}p}{(2\pi)^{3}} \left\lbrace
    1+\frac{px(px+\omega_{\vec{k}})+m^{2}_{c}}{E^{c}_{\vec{p}}E^{c}_{\vec{p}+\vec{k}}}
  \right\rbrace
  \frac{\varepsilon^{2}}{\left(E^{c}_{\vec{p}+\vec{k}}+E^{c}_{\vec{p}}+\omega_{\vec{k}}\right)^{4}} \nonumber \\
  = & 0 \ ,
\end{align}
where we have taken into account that
$E^{c}_{\vec{p}+\vec{k}}+E^{c}_{\vec{p}}+\omega_{\vec{k}}$ is positive
definite in the second step. Therefore, only mass-shift contributions to
(\ref{eq:3:phot_num_den_phys}) characterized by
$\ii\Delta\Pi^{<}_{\text{T}}(\vec{k},t_{1},t_{2})$ remain. In this
context, we would like to point out that adhering to the exact sequence
of limits, i.e., taking first $t\rightarrow\infty$ and then
$\varepsilon\rightarrow0$, is indeed crucial to eliminate the
contribution from the vacuum polarization. When first taking
$\varepsilon\rightarrow0$ at some finite time and $t\rightarrow\infty$
afterwards, the contribution from (\ref{eq:3:phot_num_den_phys}) instead
turns into
\begin{equation}
 \left.\omega_{\vec{k}}\frac{\dd^{6}n_{\gamma}(t)}{\dd^{3}x\dd^{3}k}\right|_{\text{VAC}} 
  =\frac{e^{2}}{(2\pi)^{3}}\int\frac{\dd^{3}p}{(2\pi)^{3}}
    \left\lbrace
     1+\frac{px(px+\omega_{\vec{k}})+m^{2}_{c}}{E^{c}_{\vec{p}}E^{c}_{\vec{p}+\vec{k}}}
    \right\rbrace
    \frac{1}{\left(E^{c}_{\vec{p}+\vec{k}}+E^{c}_{\vec{p}}+\omega_{\vec{k}}\right)^{2}} \ ,
\end{equation}
with the integral over $\dd^{3}p$ being linearly divergent for given
photon energy, $\omega_{\vec{k}}$. In the course of our numerical
investigations, we furthermore demonstrate that the correct sequence of
limits is also essential to obtain physically reasonable results from
the mass-shift effects. Together with (\ref{eq:3:pse}) and
(\ref{eq:3:eom_prop}), expression (\ref{eq:3:phot_num_den_phys})
describes photon emission arising from the chiral mass shift at first
order in $\alpha_{e}$ but to all orders in $g$.

\section{Quark-antiquark pair production}
\begin{figure}[htb]
 \begin{center}
  \includegraphics[height=5.0cm]{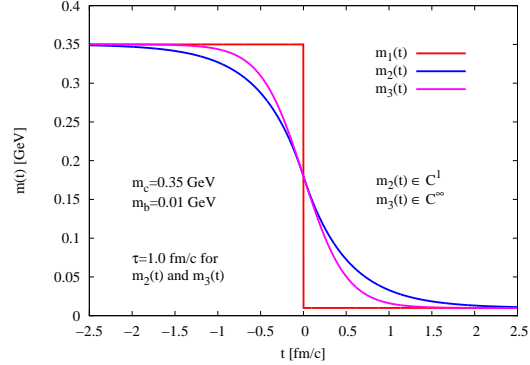}
  \caption{The change of the quark/antiquark mass from its constituent
    value, $m_{c}$, to its bare value, $m_{b}$, during the chiral phase
    transition is modeled by different mass parameterizations, $m_{i}(t)$.}
  \label{fig:4:mass}
 \end{center}
\end{figure}

Before we turn to our numerical investigations on chiral photon
production, we first provide an insertion on quark-antiquark pair
production arising from the chiral mass shift. Thereby, we consider the
pair production arising due to the chiral mass shift only. The starting point
of our investigations hence is the fermionic part of the Hamiltonian
\begin{align}
 \label{eq:4:hamiltonian}
 \hat{H}(t) & = \int \dd^{3}x \;
 \hat{\bar{\psi}}(x) \left[-\ii
   \vec{\gamma}\cdot\vec{\nabla} + m(t) \right]
 \hat{\psi}(x) \ ,
\end{align}
with the mass function, $m(t)$, given by (\ref{eq:3:mass_timedep}). As
the next step, we expand the fermion-field operator, $\hat{\psi}(x)$, in
terms of the positive- and negative-energy wavefunctions,
$\psi_{\vec{p},s\uparrow\downarrow}(t)$, i.e.,
\begin{equation}
 \label{eq:4:expand}
 \hat{\psi}(x) = \sum_{s}\int\frac{\dd^{3}p}{(2\pi)^{3}}
 \left[\psi_{\vec{p},s,\uparrow}(t)\hat{b}_{\vec{p},s}+
   \psi_{\vec{p},s,\downarrow}(t)\hat{d}^{\dagger}_{-\vec{p},s}
 \right]\ee^{\ii\vec{p}\cdot\vec{x}} \ ,
\end{equation}
with $\hat{b}_{\vec{p},s}$ and $\hat{d}_{-\vec{p},s}$ both annihilating
the initial fermionic vacuum state, $\left|0_{q\bar{q}}\right\rangle$.
Upon insertion of (\ref{eq:4:expand}) into (\ref{eq:4:hamiltonian}) we
obtain
\begin{align}
 \label{eq:4:hamiltonian_re}
 \hat{H}(t) = \sum_{s}\int\frac{\dd^{3}p}{(2\pi)^{3}}
 \left\lbrace\Omega(t)\left[
     \hat{b}^{\dagger}_{\vec{p},s}\hat{b}_{\vec{p},s}-
     \hat{d}_{-\vec{p},s}\hat{d}^{\dagger}_{-\vec{p},s} \right]+
   \Lambda(t)\hat{b}^{\dagger}_{\vec{p},s}\hat{d}^{\dagger}_{-\vec{p},s}+
   \Lambda^{*}(t)\hat{d}_{-\vec{p},s}\hat{b}_{\vec{p},s} \right\rbrace \ .
\end{align}
In order to keep the notation short, we have introduced
\begin{subequations}
 \begin{align}
   \Omega(t) & = \bar{\psi}_{\vec{p},s,\uparrow}(t)\left
     [\gamma^{i}p_{i}+m(t)\right ]\psi_{\vec{p},s,\uparrow}(t)
   =  -\bar{\psi}_{\vec{p},s,\downarrow}(t)\left [\gamma^{i}p_{i}+m(t)\right]\psi_{\vec{p},s,\downarrow}(t) \ ,   \\
   \Lambda(t) & = \bar{\psi}_{\vec{p},s,\uparrow}(t)\left
     [\gamma^{i}p_{i}+m(t)\right]\psi_{\vec{p},s,\downarrow}(t) \ .
 \end{align}
\end{subequations}
The particle number-density of quarks and antiquarks, which corresponds
to the number of produced quark-antiquark pairs per unit volume, is
extracted from (\ref{eq:4:hamiltonian_re}) by diagonalizing this
expression via a Bogolyubov transformation
\begin{subequations}
 \label{eq:4:transformation}
 \begin{eqnarray}
   \hat{\tilde{b}}_{\vec{p},s}(t)            & = &
   \xi_{\vec{p},s}(t)\hat{b}_{\vec{p},s}+\eta_{\vec{p},s}(t)\hat{d}^{\dagger}_{-\vec{p},s}
   \ , \\
   \hat{\tilde{d}}^{\dagger}_{-\vec{p},s}(t) & = &
   \xi^{*}_{\vec{p},s}(t)\hat{d}^{\dagger}_{-\vec{p},s}-\eta^{*}_{\vec{p},s}(t)\hat{b}_{\vec{p},s}
   \ .
 \end{eqnarray}
\end{subequations}
As discussed in greater detail in \cite{Michler:2012mg}, this procedure
corresponds to a re-expansion of the fermion-field operator
(\ref{eq:4:expand}) in terms of the instantaneous eigenstates of the
Hamilton density operator
$$
 \hat{h}_{\text{D}}(t) = -\ii\vec{\gamma}\cdot\vec{\nabla}+m(t) \ .
$$
As a consequence, the Hamiltonian (\ref{eq:4:hamiltonian}) reads in
terms of the transformed creation and annihilation operators
(\ref{eq:4:transformation}):
\begin{align}
 \hat{H}(t) =           & \sum_{s}\int\frac{\dd^{3}p}{(2\pi)^{3}}E_{\vec{p}}(t)
                           \left[
                            \hat{\tilde{b}}^{\dagger}_{\vec{p},s}(t)\hat{\tilde{b}}_{\vec{p},s}(t)-
                            \hat{\tilde{d}}_{-\vec{p},s}(t)\hat{\tilde{d}}^{\dagger}_{-\vec{p},s}(t)
                           \right] \nonumber \\   
            \rightarrow & \sum_{s}\int\frac{\dd^{3}p}{(2\pi)^{3}}E_{\vec{p}}(t)
                           \left[
                            \hat{\tilde{b}}^{\dagger}_{\vec{p},s}(t)\hat{\tilde{b}}_{\vec{p},s}(t)+
                            \hat{\tilde{d}}^{\dagger}_{-\vec{p},s}(t)\hat{\tilde{d}}_{-\vec{p},s}(t)
                           \right] \ .
\end{align}
In the second step, $\hat{H}(t)$ has been normal-ordered with respect to
(\ref{eq:4:transformation}) as to avoid an infinite
negative vacuum energy. Moreover, we have introduced the dispersion
relation
\begin{equation}
 \label{eq:4:dispersion}
 E_{\vec{p}}(t)=\sqrt{\vec{p}^{2}+m^{2}(t)} \ .
\end{equation}
The Bogolyubov particle number-density and thus the number of produced
quark/antiquark pairs is then defined as \cite{FW:2002}:
\begin{align}
 \label{eq:4:particle_number}
 \frac{\dd^{6}n_{q\bar{q}}(t)}{\dd^{3}x\dd^{3}p} & =
 \frac{1}{(2\pi)^{3}V}\sum_{s} \left\langle0_{q\bar{q}}\right|
 \hat{\tilde{b}}^{\dagger}_{\vec{p},s}(t)\hat{\tilde{b}}_{\vec{p},s}(t)
 \left|0_{q\bar{q}}\right\rangle \nonumber \\
 & = \frac{1}{(2\pi)^{3}V}\sum_{s} \left\langle0_{q\bar{q}}\right|
 \hat{\tilde{d}}^{\dagger}_{-\vec{p},s}(t)\hat{\tilde{d}}_{-\vec{p},s}(t)
 \left|0_{q\bar{q}}\right\rangle \nonumber \\
 & = \frac{1}{(2\pi)^{3}}\sum_{s}\left|\eta_{\vec{p},s}(t)\right|^{2} \
 ,
\end{align}
For our numerical investigations on the number of produced
quark-antiquark pairs, we consider different mass parameterizations,
which are depicted in Figure \ref{fig:4:mass}.

\begin{figure}[htb]
 \begin{center}
  \includegraphics[height=5.0cm]{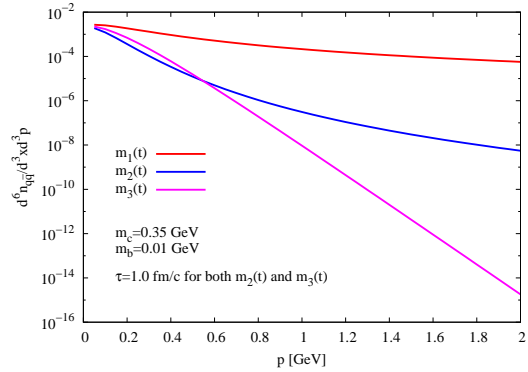}
  \caption{The decay behavior of the asymptotic particle spectra shows a
    strong sensitivity on the `smoothness' of considered mass
    parametrization, $m_{i}(t)$.}
  \label{fig:4:chiral_asymptcomp}
 \end{center}
\end{figure}

Thereby, $m_{1}(t)$ describes an instantaneous change from the
constituent mass, $m_{c}$, to the bare mass, $m_{b}$, at $t=0$, whereas
$m_{2}(t)$ and $m_{3}(t)$ each describe a change over a finite interval
of time. The latter two mass parameterizations differ with respect to their
order of differentiability. $m_{2}(t)$ is continuously differentiable
once, whereas $m_{3}(t)$ is continuously differentiable infinitely many
times. For both parameterizations, we assume a transition time of
$\tau=1 \; \fm/c$. Figure \ref{fig:4:chiral_asymptcomp} compares the
asymptotic particle spectra for the different mass parameterizations,
$m_{i}(t)$.

For $m_{1}(t)$, the quark/antiquark occupation numbers scale as
$1/p^{2}$ for $p\gg m_{c},m_{b}$, which implies that the total number
density and the total energy density of the produced quarks and
antiquarks are linearly and quadratically divergent,
respectively. However, this artifact is cured when turning from
instantaneous mass change to a mass change over a finite interval of
time. In particular, the quark/antiquark occupation numbers are
suppressed to $1/p^{6}$ for $m_{2}(t)$ such that both the total number
density and the total energy density are rendered finite. Moreover, if
one turns from $m_{2}(t)$ to $m_{3}(t)$, which is continuously
differentiable infinitely many times and hence describes the most
physical scenario, the quark/antiquark occupation numbers are suppressed
even further to an exponential decay. One encounters the same
sensitivity on the `smoothness' of the mass parameterization if the mass
is not only changed from its constituent value, $m_{c}$, to its bare
value, $m_{b}$, but also back to its constituent value after a certain
interval of time as to take into account the finite lifetime of the
chirally restored phase \cite{Michler:2012mg}.

\begin{figure}[htb]
 \center
 \includegraphics[height=5.0cm]{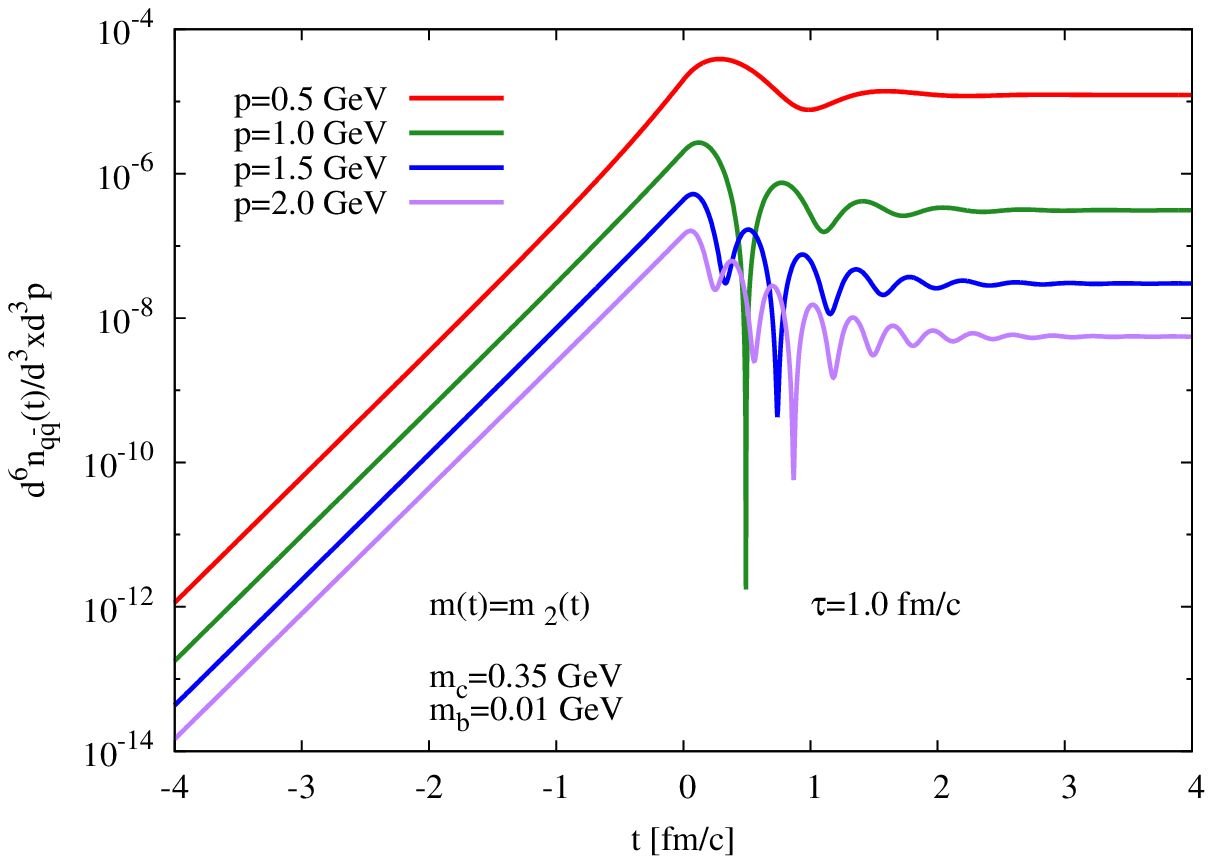}
 \includegraphics[height=5.0cm]{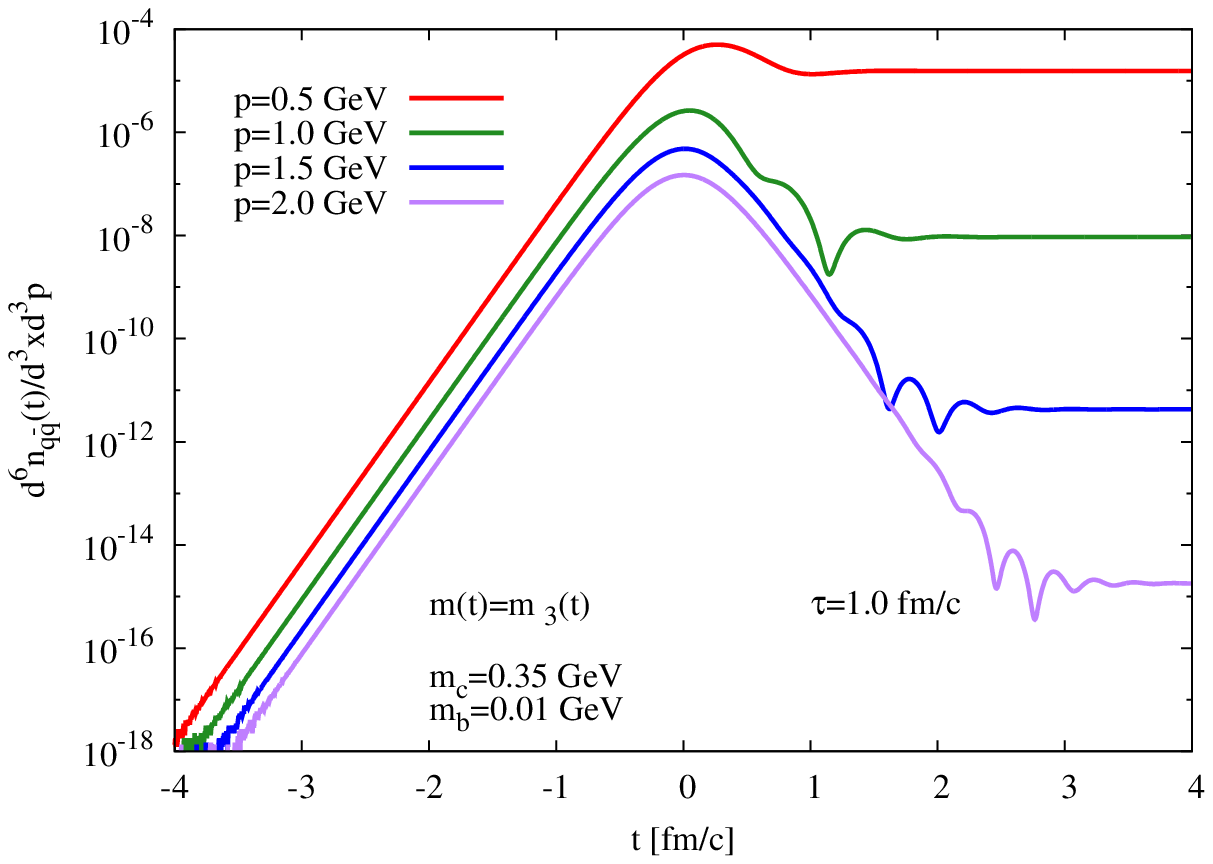}
 \caption{Time dependence of the quark/antiquark occupation numbers for
   $m_{2}(t)$ (left panel) and $m_{3}(t)$ (right panel) for $\tau=1.0$
   fm/$c$. For large values of $p$, they exhibit an `overshoot' over
   their asymptotic value around $t=0$, which is particularly
   distinctive for $m_{3}(t)$.}
 \label{fig:4:chiral_timedep}
\end{figure}

In addition to the asymptotic particle spectra, we also consider the
time evolution of the quark/antiquark occupation numbers, which is shown
in Figure \ref{fig:4:chiral_timedep}.

As one can see, the quark/antiquark occupation numbers exhibit a strong
`overshoot' over their asymptotic value in the region of strong mass
gradients, which is especially distinctive for $m_{3}(t)$. In
particular, the quark/antiquark occupation numbers scale as $1/p^{4}$
for large $p$ when the mass is being changed from $m_{c}$ to
$m_{b}$. Such a scaling behavior means that the total number density is
finite, whereas the total energy density is divergent. This divergence,
however, only shows up in the region of strong mass gradients and
disappears again as soon as the quark/antiquark mass has reached its
final bare value, $m_{b}$.

At first sight, the temporary logarithmic divergence might be
disturbing. In this context, it is, however, important to point out that
only the asymptotic energy density, i.e., for $t\rightarrow\infty$
constitutes a well-defined physical quantity since the interpretation of
(\ref{eq:4:particle_number}) as a particle number-density is only
justified for asymptotic times where $\dot{m}(t)=0$. The reason is that
the dispersion relation (\ref{eq:4:dispersion}) then actually
characterizes free and thus detectable particles, wheres it describes
quasiparticles for $\dot{m}(t)\neq0$. In particular, the asymptotic
energy density is finite as long as the mass parameterization, $m(t)$,
is chosen smooth enough, which represents a physically reasonable
condition. Moreover, the temporary logarithmic divergence in the energy
density does not manifest itself in form of a similar pathology neither
in the total number density nor in the total energy density of the
emitted photons.

\section{Photon production}
Our investigations on pair production have shown that the asymptotic
quark/antiquark occupation numbers exhibit a very strong sensitivity on
the `smoothness' of the mass parameterization, $m(t)$. In particular,
the asymptotic particle spectra are rendered UV integrable if the mass
change is assumed to take place over a finite interval of time,
$\tau$. We now determine whether the asymptotic photon number-density
exhibits a similar sensitivity. In this context, we recall that only
asymptotic quantities constitute observables whereas the corresponding
expressions at finite times do, in general, not allow for a similar
interpretation.

For our investigations on chiral photon production, we again consider
the different mass parameterizations shown in Figure
\ref{fig:4:mass}. For the special case of an instantaneous mass shift,
the loop integral entering (\ref{eq:3:phot_num_den_phys}) features a
linear divergence. This divergence arises from the Bogolyubov particle
numbers $\propto 1/p^{2}$ for large $p$ \cite{Michler:2012mg}. This
behavior is an artifact from the instantaneous mass shift and is removed
if the mass shift is assumed to take place over a finite interval of
time. Hence, from the conceptional point of view this divergence does
not require a specific kind of renormalization and is regulated by a
cutoff at $p=\Lambda_{\text{C}}$. Figure 7 shows the resulting photon
spectra for the different mass parameterizations. Thereby, we have chosen
$\Lambda_{\text{C}}=20$ GeV for $m(t)=m_{1}(t)$.
\begin{figure}[htb]
 \begin{center}
  \includegraphics[height=5.0cm]{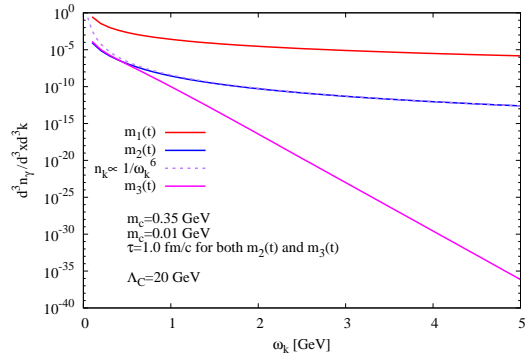}
  \caption{The decay behavior of the asymptotic photon spectra in the UV
    domain is highly sensitive to the `smoothness' of the mass
    parameterization. In particular, they are rendered integrable in
    this domain if the change of the quark/antiquark mass is assumed to
    take place over a finite interval of time, $\tau$.}
  \label{fig:5:photspec_comp}
 \end{center}
\end{figure}

We see that for an instantaneous mass change, the photon number-density
scales as $1/\omega^{3}_{\vec{k}}$ for large $\omega_{\vec{k}}$, which
means that for given given $\Lambda_{\text{C}}$, the total number
density and the total energy density of the emitted photons are
logarithmically and linearly divergent, respectively. In contrast to our
earlier model approach \cite{Michler:2009hi}, this artifact is now fully
removed if we turn from an instantaneous mass shift to a mass shift over
a finite interval of time. For $m_{2}(t)$, which is continuously
differentiable once, the photon number-density is suppressed from
$1/\omega^{3}_{\vec{k}}$ to $1/\omega^{6}_{\vec{k}}$ in the UV domain
such that both the total number density and the total energy density of
the emitted photons are UV finite. Moreover, if one turns from
$m_{2}(t)$ to $m_{3}(t)$, which continuously differentiable infinitely
many times, the photon number-density is suppressed even further to an
exponential decay. The sensitivity of this quantity to the `smoothness'
of the mass parameterization is the same if the quark/antiquark mass is
changed back to its constituent value, $m_{c}$, after a certain period
of time as to take into account the finite lifetime of the chirally
restored phase \cite{Michler:2012mg}.

\begin{figure}[htb]
 \begin{center}
  \includegraphics[height=5.0cm]{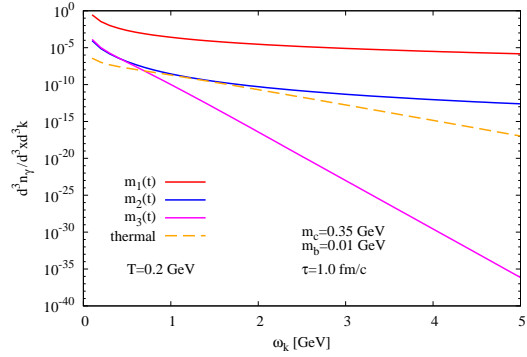}
  \caption{Comparison of first-order mass-shift contributions to
    leading-order thermal contributions. For $i=3$ most likely
    characterizing a physical scenario, the contributions arising from
    the chiral mass shift are clearly subdominant for
    $\omega_{\vec{k}}\gtrsim 1 \; \GeV$.}
  \label{fig:5:thermcomp}
 \end{center}
\end{figure}

As the artifact with the UV behavior encountered in
\cite{Wang:2000pv,Wang:2001xh,Boyanovsky:2003qm} and still partly in
\cite{Michler:2009hi} is now resolved, it is convenient to compare our
results to leading-order thermal contributions. This is done in Figure
\ref{fig:5:thermcomp}. Thereby, the thermal contributions have been
obtained by integrating the leading-order thermal rate taken from
\cite{Arnold:2001ms} over the expected lifetime of the chirally restored
phase of $\tau_{\text{L}}=4.0$ fm/c at a temperature of $T=0.2$ GeV. In
this context, we note again that first-order photon production does not
occur in thermal equilibrium. There the leading-order contributions are
of linear order both in the electromagnetic coupling constant,
$\alpha_{e}$, and the the strong coupling constant $\alpha_{s}$, and
hence of overall second order.

Nevertheless, we see that chiral photon production does not generally
dominate over leading-order thermal photon production. For
$m(t)=m_{3}(t)$, which describes the most physical scenario as it is
continuously differentiable infinitely many times, we see that chiral
photon production is even subdominant by several orders of magnitude in
the UV domain.

\begin{figure}[htb]
 \begin{center}
  \includegraphics[height=5.0cm]{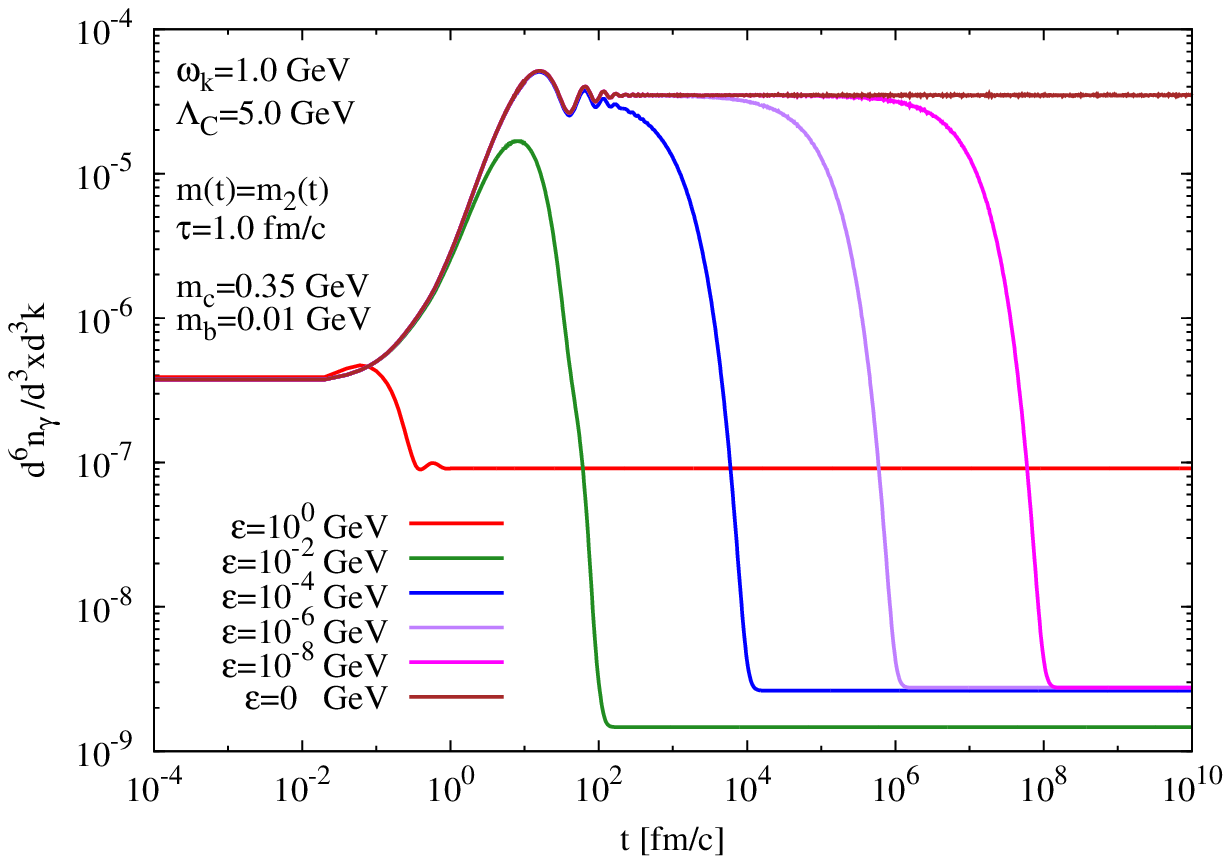}
  \includegraphics[height=5.0cm]{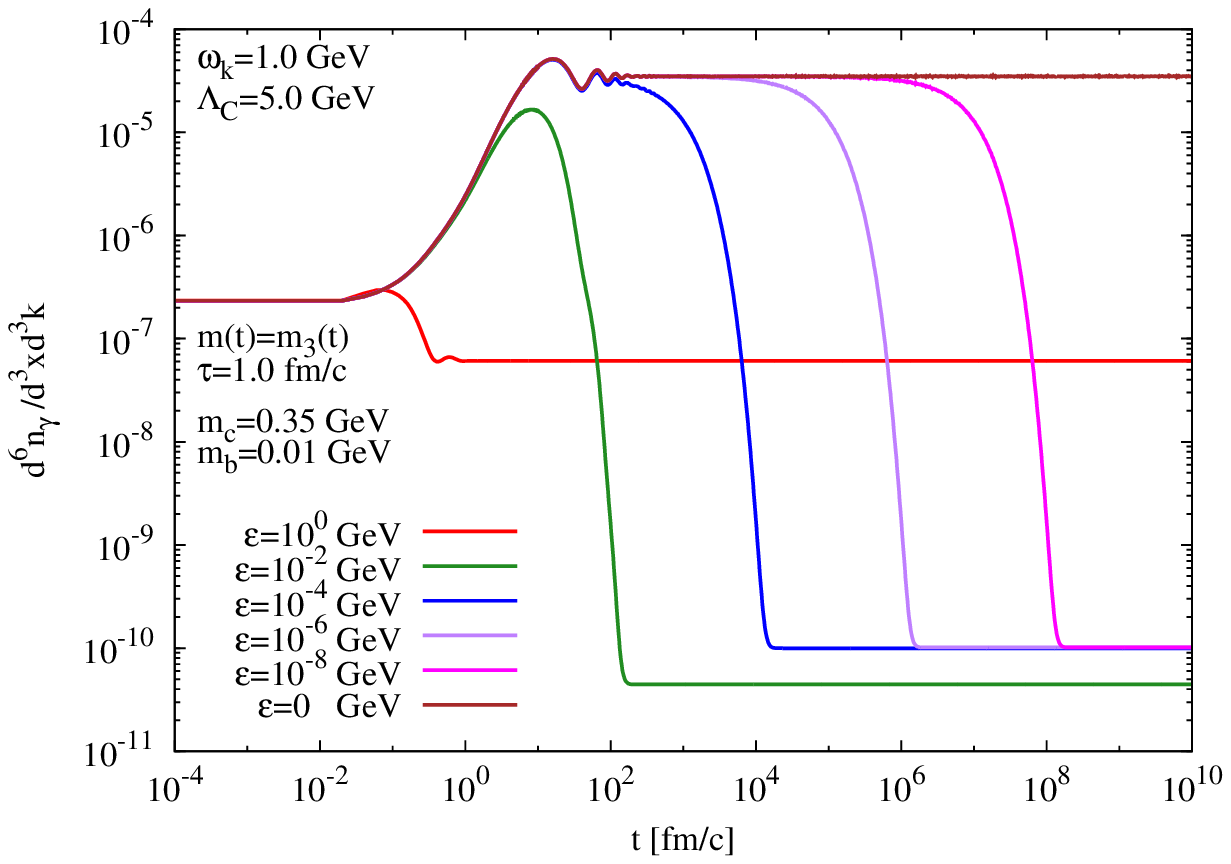}
  \caption{Time evolution of the mass-shift contribution for different
    values of $\varepsilon$ for $m_{2}(t)$ (left panel) and $m_{3}(t)$
    (right panel).}
  \label{fig:5:evolution}
 \end{center}
\end{figure}

For completeness, we would still like to demonstrate that adhering to
the correct sequence of limits leading to
(\ref{eq:3:phot_num_den_phys}), i.e., taking \emph{first}
$t\rightarrow\infty$ and \emph{then} $\varepsilon\rightarrow0$ is
crucial not only to eliminate a possible unphysical contribution from
the vacuum polarization, but also to obtain physically reasonable
results from mass-shift effects. For this purpose, we consider the time
evolution of the pure mass-shift contribution to
(\ref{eq:3:phot_num_den_phys}) for different values of $\varepsilon$
which is shown in Figure \ref{fig:5:evolution}.

Taking first $\epsilon\rightarrow0$ at some finite time corresponds to
the curve labeled by $\epsilon=0$ in each case. If we then take the
subsequent limit $t\rightarrow\infty$, we obtain an asymptotic value for
the mass-shift contribution which is by several orders of magnitudes
larger than the value against which the asymptotic values for finite
$\epsilon$ converge in the limit $\varepsilon\rightarrow0$ and which
accordingly corresponds to (\ref{eq:3:phot_num_den_phys}). The reason
why the asymptotic value obtained for an interchanged sequence of limits
must be unphysical is that it coincides with the transient value of the
mass-shift contribution for sufficiently small values of $\varepsilon$. On 
the other hand, expression (\ref{eq:3:photnum_prelim}) does not allow for 
an interpretation as a photon number-density at finite times since one either 
has no free asymptotic states or one has to introduce an artificial switching-off 
of the electromagnetic interaction at the point of time at which (\ref{eq:3:photnum_prelim}) 
is considered. By means of the latter procedure, a violation of the Ward-Takahashi identities 
furthermore comes into play again \cite{Michler:2012mg}. 

The same conceptual problem arises when only using an adiabatic switching-on of 
the electromagnetic interaction for $t\rightarrow-\infty$ but no switching-off for $t\rightarrow\infty$. 
Such a procedure has been introduced in \cite{Serreau:2003wr} as to describe intial correlations 
at some $t=t_{0}$ evolving from an uncorrelated initial state at $t\rightarrow-\infty$. In fact, the 
excess of the mass-shift contribution over its asymptotic value at finite times arises from spurious 
transient contributions that again lead to an unphysical UV scaling behavior and must be eliminated 
by switching the electromagnetic interaction off again according to (\ref{eq:3:adiabatic}). 

In addition to the vacuum contribution and the pure mass-shift
contribution, the photon self-energy (\ref{eq:3:pse}) also contains a
contribution which describes the interference between the vacuum and the
mass-shift effects. This contribution is also eliminated by the sequence
of limits leading to (\ref{eq:3:phot_num_den_phys}).  A more detailed
consideration can be found in \cite{Michler:2012mg}.

\section{Summary, conclusions and outlook}
In this work, we have investigated photon emission during the chiral
phase transition in the early stage of a heavy-ion collision. During
this phase transition, the quark mass is changed from its constituent
value, $m_{c}$, to its bare value, $m_{b}$, which leads to the
spontaneous and non-perturbative pair production of quarks and
antiquarks \cite{Greiner:1995ac,Greiner:1996wz}. This effectively
contributes to the creation of the QGP, and we have investigated the
photon emission arising from this pair-creation process. In particular,
our investigations are relevant in the context of the question how the
finite lifetime of the QGP during a heavy-ion collision affects the
photon emission from it.

Earlier investigations on this topic had been accompanied by two
artifacts. In the first place, the photon number-density contained a
divergent contribution from the vacuum polarization. Furthermore, the
remaining contributions to this quantity were not integrable in
the UV domain. In particular, the total number density and the total
energy density of the emitted photons were logarithmically and linearly
divergent, respectively. It has been our original motivation to resolve
these conceptional problems in a satisfactory manner.

For this purpose, we have first pursued a model approach
\cite{Michler:2009hi} where the finite lifetime of the QGP is simulated
by time-dependent quark/antiquark occupation numbers in the photon
self-energy. By means of this procedure, we have been able to
renormalize the divergent contribution from the vacuum polarization in a
consistent manner, but it has not been possible to solve the remaining
problem with the UV behavior to full extent. We have assumed that this
remaining shortcoming results from a violation of the Ward-Takahashi
identities. This aspect has been our eventual motivation to consider
photon production arising from the change of the quark/antiquark mass
during the chiral phase transition since this scenario allows for a
first-principle description by which the Ward-Takahashi identities are
conserved.

In particular, the change of the quark/antiquark mass has been modeled
by a Yukawa-like source term in the QED Lagrangian which couples the
quarks and antiquarks to a scalar and time-dependent background
field. As in
\cite{Wang:2000pv,Wang:2001xh,Boyanovsky:2003qm,Michler:2009hi}, we have
restricted ourselves to first-order QED processes. These processes are
kinematically possible since the background field acts as a source of
additional energy. On the other hand, the coupling to the background
field has been resummed to all orders as to properly take into account
the non-perturbative nature of the pair-creation process.

In contrast to
\cite{Wang:2000pv,Wang:2001xh,Boyanovsky:2003qm,Michler:2009hi}, we have
not considered the photon number-density at finite times but for free
asymptotic states. Such states have been obtained by switching the
electromagnetic interaction according to the Gell-Mann and Low
theorem. The photon number-density has then been considered in the
asymptotic time limit $t\rightarrow\infty$ and we have taken the
switching parameter to zero at the very end of our calculation. By means
of this procedure, the photon number-density does not contain any
unphysical contributions from the vacuum polarization and is furthermore
rendered UV integrable for physically reasonable mass
parameterizations. 

In particular, we have shown that the consideration
of this quantity for free asymptotic states is indeed crucial to obtain
such physically reasonable results. The reason is that a 
consistent definition actually is only possible 
for free asymptotic states, whereas a similar interpretation of the respective 
formal expression is usually not justified at finite times, $t$, since one then 
does not have free asymptotic states. The same problem occurs if the 
electromagnetic interaction is only switched on adiabatically from $t\rightarrow-\infty$ 
but not switched off again for $t\rightarrow\infty$. This procedure has been suggested in \cite{Serreau:2003wr} 
as to implement initial correlations at some $t=t_{0}$ evolving from an uncorrelated 
initial state at $t\rightarrow-\infty$.

Moreover, the consideration of the `photon number-density' at finite 
times effectively comes along with a violation of the Ward-Takahashi identities. Our 
investigations hence support the respective concern raised in \cite{Fraga:2003sn,Fraga:2004ur} 
towards \cite{Wang:2000pv,Wang:2001xh,Boyanovsky:2003qm}.

\begin{figure}[htb]
 \begin{center}
  \includegraphics[height=5.0cm]{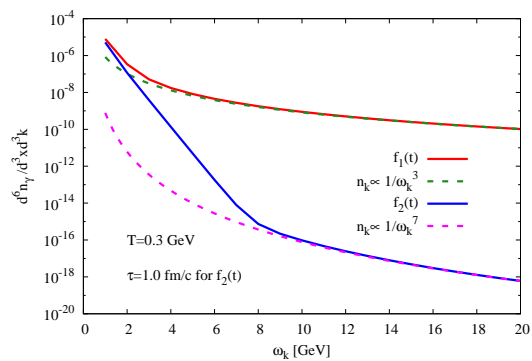}
  \caption{The photon spectra are also rendered UV integrable within our
    model description \cite{Michler:2009hi} if they are considered for
    free asymptotic states and if the time evolution of the QGP is
    modeled in a suitable manner, i.e., if its formation is assumed to
    take place over a finite interval of time.}
  \label{fig:6:model_aspt}
 \end{center}
\end{figure}

Consequently, our results indicate that the remaining artifact with the UV
behavior within our model description \cite{Michler:2009hi} results from
an inconsistent definition of the photon number-density at finite
times. This in turn imposes the question whether this artifact is also
removed if the photon number-density is instead considered for free
asymptotic states, which are obtained by switching the electromagnetic 
interaction according to the Gell-Mann and Low theorem, i.e.,
\begin{equation}
 \label{eq:5:adiabatic}
 \hat{H}_{\text{EM}}(t) = e\int\dd^{3}x\mbox{ }\hat{\bar{\psi}}(x)\gamma_{\mu}\hat{\psi}(x)\hat{A}^{\mu}(x)
                          \rightarrow
                          \ee^{-\varepsilon|t|}\hat{H}_{\text{EM}}(t) \
                          , \quad\text{with}\quad \varepsilon > 0 \ .
\end{equation}
An accordant revision of \cite{Michler:2009hi}, for which the resulting photon spectra 
are depicted in Figure \ref{fig:6:model_aspt}, shows that this is indeed the case.

As in \cite{Wang:2000pv,Wang:2001xh,Boyanovsky:2003qm}, the photon
number-density decays as $1/\omega^{3}_{\vec{k}}$ in the UV domain if
the quark/antiquark occupation numbers are switched on
instantaneously. If we turn from an instantaneous switching to a
switching over a finite interval of time, however, the photon
number-density is suppressed from $1/\omega^{3}_{\vec{k}}$ to
$1/\omega^{7}_{\vec{k}}$ which means that the total number density and
the total energy density of the radiated photons are both rendered UV
finite. To the contrary, the UV scaling behavior is solely suppressed 
from $1/\omega^{3}_{\vec{k}}$ to $1/\omega^{3.8}_{\vec{k}}$ when turning from $f_{1}(t)$ to $f_{2}(t)$ if the 
'photon number-density' is instead considered at finite times (see Figure \ref{fig:2:photspec}). 
Our outlook to future investigations is hence as follows.

In the first place, the actual role of the Ward-Takahashi identities,
which are violated in \cite{Michler:2009hi} but conserved within our
first-principle approach to chiral photon production requires deeper
consideration. In particular, the fact that our model description
\cite{Michler:2009hi} leads to UV integrable photon spectra despite the
violation of these identities does not necessarily disprove our earlier
conjecture concerning their role. The reason is that the Ward-Takahashi
identities can be violated in two different ways:
\begin{itemize}
\item On the one hand, they can be violated directly by making ad-hoc
  assumptions on the two-time dependence of the photon self-energy,
  which has been the case in \cite{Michler:2009hi}.
\item On the other hand, they can also be violated indirectly by
  considering the `photon number-density' at finite times even though
  they are formally fulfilled at first. This has been shown in
  \cite{Michler:2012mg}.
\end{itemize}
Since our model description \cite{Michler:2009hi} leads to a
UV integrable photon number-density if this quantity is considered for
free asymptotic states, we hence have to determine if and, as the case
may be, why only an indirect violation of the Ward-Takahashi identities,
which occurs in addition to the direct violation when considering a
transient `photon number-density' within \cite{Michler:2009hi}, leads to
artificial results.

Moreover, it is of particular interest whether our previous asymptotic 
description allows for a consistent extension to finite times and/or which 
alternative quantities can be considered to describe the time evolution of 
the electromagnetic sector during a heavy-ion collision.

\bibliographystyle{JHEP}

\begin{thebibliography}{10}

\bibitem{BS:1992}
G.~S. Bali and K.~Schilling, {\it Static quark-antiquark potential: Scaling
  behaviour and finite size effects in su(3) lattice gauge theory},  {\em Phys.
  Rev. D} {\bf 46} (1992) 2626--2646.

\bibitem{GW:1973}
D.~Gross and F.~Wilczek, {\it Ultraviolet behaviour of non-abelian gauge
  theories},  {\em Phys. Rev. Lett.} {\bf 30} (1973) 1343--1346.

\bibitem{Pol:1973}
H.~D. Politzer, {\it Reliable perturbative results for strong interactions},
  {\em Phys. Rev. Lett.} {\bf 30} (1973) 1346--1349.

\bibitem{Shuryak:1978ij}
E.~V. Shuryak, {\it {Quark-Gluon Plasma and Hadronic Production of Leptons,
  Photons and Psions}},  {\em Phys. Lett. B} {\bf 78} (1978) 150.

\bibitem{Shuryak:1977ut}
E.~V. Shuryak, {\it {Theory of Hadronic Plasma}},  {\em Sov. Phys. JETP} {\bf
  47} (1978) 212--219.

\bibitem{Yag:2005}
K.~Yagi, T.~Hatsuda, and Y.~Miake, {\it {Quark-gluon plasma: From big bang to
  little bang}},  {\em Camb. Monogr. Part. Phys. Nucl. Phys. Cosmol.} {\bf 23}
  (2005) 1--446.

\bibitem{Muller:2006ee}
B.~M{\"u}ller and J.~L. Nagle, {\it {Results from the relativistic heavy ion
  collider}},  {\em Ann. Rev. Nucl. Part. Sci.} {\bf 56} (2006) 93--135.

\bibitem{Friman:2011zz}
B.~Friman et~al., {\it {The CBM physics book: Compressed baryonic matter in
  laboratory experiments}},  {\em Lect. Notes Phys.} {\bf 814} (2011) 1--980.

\bibitem{Schw61}
J.~S. Schwinger, {\it Brownian motion of a quantum oscillator},  {\em J. Math.
  Phys.} {\bf 2} (1961) 407--432.

\bibitem{Bakshi:1962dv}
P.~M. Bakshi and K.~T. Mahanthappa, {\it {Expectation value formalism in
  quantum field theory. 1.}},  {\em J. Math. Phys.} {\bf 4} (1963) 1--11.

\bibitem{Bakshi:1963bn}
P.~M. Bakshi and K.~T. Mahanthappa, {\it {Expectation value formalism in
  quantum field theory. 2.}},  {\em J. Math. Phys.} {\bf 4} (1963) 12--16.

\bibitem{Kel64}
L.~Keldysh, {\it {Diagram} {Technique} for {Nonequilibrium} {Processes}},  {\em
  Zh. Eks. Theor. Fiz.} {\bf 47} (1964) 1515. [Sov. Phys. JETP \textbf{20},
  1018 (1965)].

\bibitem{Cra68}
R.~Craig, {\it {Perturbation} {Expansion} for {Real-Time} {Green's}
  {Functions}},  {\em J. Math. Phys.} {\bf 9} (1968) 605.

\bibitem{Danielewicz:1982kk}
P.~Danielewicz, {\it {QUANTUM THEORY OF NONEQUILIBRIUM PROCESSES. I}},  {\em
  Ann. Phys.} {\bf 152} (1984) 239--304.

\bibitem{Chou:1984es}
K.-c. Chou, Z.-b. Su, B.-l. Hao, and L.~Yu, {\it {Equilibrium and
  Nonequilibrium Formalisms Made Unified}},  {\em Phys. Rept.} {\bf 118} (1985)
  1.

\bibitem{Landsman:1986uw}
N.~Landsman and C.~van Weert, {\it {Real and Imaginary Time Field Theory at
  Finite Temperature and Density}},  {\em Phys. Rept.} {\bf 145} (1987) 141.

\bibitem{Greiner:1998vd}
C.~Greiner and S.~Leupold, {\it {Stochastic interpretation of Kadanoff-Baym
  equations and their relation to Langevin processes}},  {\em Ann. Phys.} {\bf
  270} (1998) 328--390.

\bibitem{Berges:2001fi}
J.~Berges, {\it {Controlled nonperturbative dynamics of quantum fields
  out-of-equilibrium}},  {\em Nucl. Phys. A} {\bf 699} (2002) 847--886.

\bibitem{Nahrgang:2011mg}
M.~Nahrgang, S.~Leupold, C.~Herold, and M.~Bleicher, {\it {Nonequilibrium
  chiral fluid dynamics including dissipation and noise}},  {\em Phys. Rev. C}
  {\bf 84} (2011) 024912.

\bibitem{Nahrgang:2011mv}
M.~Nahrgang, S.~Leupold, and M.~Bleicher, {\it {Equilibration and relaxation
  times at the chiral phase transition including reheating}},  {\em Phys. Lett.
  B} {\bf 711} (2012) 109--116.

\bibitem{P1984305}
P.~Danielewicz, {\it Quantum theory of nonequilibrium processes ii. application
  to nuclear collisions},  {\em Ann. Phys.} {\bf 152} (1984), no.~2 305 -- 326.

\bibitem{Greiner:1994xm}
C.~Greiner, K.~Wagner, and P.~G. Reinhard, {\it {Memory effects in relativistic
  heavy ion collisions}},  {\em Phys. Rev. C} {\bf 49} (1994) 1693--1701.

\bibitem{Kohler:1995zz}
H.~S. Kohler, {\it {Memory and correlation effects in nuclear collisions}},
  {\em Phys. Rev. C} {\bf 51} (1995) 3232--3239.

\bibitem{Xu:1999aq}
Z.~Xu and C.~Greiner, {\it {Stochastic treatment of disoriented chiral
  condensates within a Langevin description}},  {\em Phys. Rev. D} {\bf 62}
  (2000) 036012.

\bibitem{Juchem:2004cs}
S.~Juchem, W.~Cassing, and C.~Greiner, {\it {Nonequilibrium quantum-field
  dynamics and off-shell transport for $\phi^4$-theory in 2+1 dimensions}},
  {\em Nucl. Phys. A} {\bf 743} (2004) 92--126.

\bibitem{Juchem:2003bi}
S.~Juchem, W.~Cassing, and C.~Greiner, {\it {Quantum dynamics and
  thermalization for out-of-equilibrium $\phi^4$-theory}},  {\em Phys. Rev. D}
  {\bf 69} (2004) 025006.

\bibitem{Schenke:2005ry}
B.~Schenke and C.~Greiner, {\it {Dilepton production from hot hadronic matter
  in nonequilibrium}},  {\em Phys. Rev. C} {\bf 73} (2006) 034909.

\bibitem{Schenke:2006uh}
B.~Schenke and C.~Greiner, {\it Dilepton yields from brown-rho scaled vector
  mesons including memory effects},  {\em Phys. Rev. Lett.} {\bf 98} (2007)
  022301.

\bibitem{Michler:2009dy}
F.~Michler, B.~Schenke, and C.~Greiner, {\it {Memory effects in radiative jet
  energy loss}},  {\em Phys. Rev. D} {\bf 80} (2009) 045011.

\bibitem{Wang:2000pv}
S.-Y. Wang and D.~Boyanovsky, {\it {Enhanced photon production from quark-gluon
  plasma: Finite-lifetime effect}},  {\em Phys. Rev. D} {\bf 63} (2001) 051702.

\bibitem{Wang:2001xh}
S.-Y. Wang, D.~Boyanovsky, and K.-W. Ng, {\it {Direct photons: A nonequilibrium
  signal of the expanding quark-gluon plasma at RHIC energies}},  {\em Nucl.
  Phys. A} {\bf 699} (2002) 819--846.

\bibitem{Boyanovsky:2003qm}
D.~Boyanovsky and H.~J. de~Vega, {\it {Are direct photons a clean signal of a
  thermalized quark gluon plasma?}},  {\em Phys. Rev. D} {\bf 68} (2003)
  065018.

\bibitem{Fraga:2003sn}
E.~Fraga, F.~Gelis, and D.~Schiff, {\it {Remarks on transient photon production
  in heavy ion collisions}},  {\em Phys. Rev. D} {\bf 71} (2005) 085015.

\bibitem{Fraga:2004ur}
E.~S. Fraga, F.~Gelis, and D.~Schiff, {\it {Transient photon production in a
  QGP}},  {\em AIP Conf. Proc.} {\bf 739} (2005) 437--439.

\bibitem{Arleo:2004gn}
F.~Arleo et~al., {\it {Photon physics in heavy ion collisions at the LHC}},
  \href{http://xxx.lanl.gov/abs/hep-ph/0311131}{{\tt hep-ph/0311131}}.

\bibitem{Boyanovsky:2003rw}
D.~Boyanovsky and H.~J. de~Vega, {\it {Photon production from a thermalized
  quark gluon plasma: Quantum kinetics and nonperturbative aspects}},  {\em
  Nucl. Phys. A} {\bf 747} (2005) 564--608.

\bibitem{Michler:2009hi}
F.~Michler, B.~Schenke, and C.~Greiner, {\it {Finite lifetime effects on the
  photon production from a quark-gluon plasma}},  {\em Proceedings of the XLVII
  International Winter Meeting on Nuclear Physics} (2010)
  [\href{http://xxx.lanl.gov/abs/ 0906.173}{{\tt  0906.173}}].

\bibitem{Michler:2012mg}
F.~Michler, H.~van Hees, D.~D. Dietrich, S.~Leupold, and C.~Greiner, {\it
  {Off-equilibrium photon production during the chiral phase transition}},
  \href{http://xxx.lanl.gov/abs/1208.6565}{{\tt arXiv:1208.6565}}.

\bibitem{Greiner:1995ac}
C.~Greiner, {\it {Quark pair production in a rapid chiral phase transition}},
  {\em Z. Phys. A} {\bf 351} (1995) 317--324.

\bibitem{Greiner:1996wz}
C.~Greiner, {\it {Quark pair production in a rapid chiral phase transition}},
  {\em Prog. Part. Nucl. Phys.} {\bf 36} (1996) 395--404.

\bibitem{FW:2002}
A.~L. Fetter and J.~D. Walecka, {\em Quantum Theory of Many Particle Systems}.
\newblock Dover Publications, Mineola, NY, 2002.

\bibitem{Arnold:2001ms}
P.~B. Arnold, G.~D. Moore, and L.~G. Yaffe, {\it {Photon emission from quark
  gluon plasma: Complete leading order results}},  {\em JHEP} {\bf 0112} (2001)
  009.

\bibitem{Serreau:2003wr}
J.~Serreau, {\it Out-of-equilibrium electromagnetic radiation},  {\em JHEP}
  {\bf 05} (2004) 078.

\end{thebibliography}

\providecommand{\href}[2]{#2}\begingroup\raggedright\endgroup

\end{document}